 \documentclass[12pt,letter]{iopart}
\usepackage{iopams}
\usepackage{cite}
\usepackage{graphicx}   
\usepackage{hyperref}   

\newtheorem{theorem}{\bf Theorem}
\newtheorem{corollary}{\bf Corollary}
\newtheorem{lemma}[theorem]{Lemma}

\newcommand\T{\rule{0pt}{2.6ex}}       
\newcommand\B{\rule[-1.2ex]{0pt}{0pt}} 

\begin{document}
 \title{ When do knots in light stay knotted?}
\author{Hridesh Kedia$^1$, Daniel Peralta-Salas$^2$ and William T.M. Irvine$^1$ }
\address{$^1$ James Franck Institute and Department of Physics, The University of Chicago, 929 E 57th St, Chicago, IL 60637, USA}
\address{$^2$ Instituto de Ciencias Matem\'{a}ticas, Consejo Superior de Investigaciones Cient\'{i}ficas, 28049 Madrid, Spain}

\ead{hridesh@uchicago.edu}
\ead{dperalta@icmat.es}
\ead{wtmirvine@uchicago.edu}

\begin{abstract}
An initially knotted light field will stay knotted if it satisfies a set of nonlinear, geometric constraints, i.e. the null conditions, for all space-time. However, the question of when an initially null light field stays null has remained challenging to answer.  
By establishing a mapping between Maxwell's equations and transport along the flow of a pressureless Euler fluid, we show that an initially analytic null light field stays null if and only if the flow of the initial Poynting field is shear-free, giving a design rule for the construction of persistently knotted light fields. 
Furthermore we outline methods for constructing initially knotted null light fields, and initially null, shear-free light fields, and give sufficient conditions for the magnetic (or electric) field lines of a null light field to lie tangent to surfaces. Our results pave the way for the design of persistently knotted light fields and the study of their field line structure.
\end{abstract}

\section{Introduction}

The concept that physical fields can have a geometry of their own, has a rich tradition that dates back to Faraday's concept of lines of force \cite{faraday_lines_1852}. 
This concept was followed shortly thereafter with the idea that you could in principle tie knots in fields, as in  Lord Kelvin's celebrated conjecture that atoms were vortices in the aether tied into knots~\cite{kelvin_vortex_1869}.  
Such knots, have now become an experimental reality in a wide variety of systems such as the vortex lines of a fluid \cite{kleckner_creation_2013}, the topological defect lines of liquid crystals \cite{tkalec_reconfigurable_2011}, lines of darkness in optical beams \cite{dennis_isolated_2010}, and spinor Bose-Einstein condensates \cite{hall_tying_2016}. Theoretical approaches have also moved beyond speculation and stable knotted structures have been shown to exist as solutions of nonlinear field theories such as Euler flows \cite{enciso_knots_2012,enciso_existence_2015}, the Skyrme-Faddeev model \cite{faddeev_stable_1997,houghton_rational_1998,manton_topological_2004}, and the AFZ model \cite{aratyn_exact_1999}.

Within this plethora of systems, light fields offer a unique opportunity for studying knotted dynamical structures analytically. Unlike the systems mentioned above, Maxwell's equations are linear, offering the enticing possibility of finding exact expressions for evolving knotted fields as well as providing the potential means of transferring knottedness to other systems such as quantum fluids and plasmas.
It is perhaps not obvious that such a linear field theory could contain persistent knotted structures, 
however an extensively studied  example of a light field containing persistent links (the Hopfion solution) \cite{ranada_knotted_1990,ranada_topological_1989,ranada_topological_1992,irvine_linked_2008,urbantke_hopf_2003,trautman_solutions_1977,arrayas_knots_2017} showed that it is possible.
This solution was recently further generalized to an infinite family of knotted light fields \cite{kedia_tying_2013}. 

In these solutions, the lines of the magnetic (electric) field are either linked to each other, knotted, or tangled in an organized way. 
As the light fields evolve in time, the closed loops formed by magnetic (electric) field lines deform smoothly like elastic filaments embedded in a flow, preserving the knotted structures they encode.

At the heart of the evolution of these persistently knotted light fields, is  a set of geometric constraints known as the null conditions \cite{robinson_null_1961}:  
\begin{equation}
\mathbf{E}\cdot\mathbf{B}=0\, , \,\mathbf{E}\cdot\mathbf{E} = \mathbf{B}\cdot\mathbf{B} \qquad \quad \textrm{\small (c=1 assumed from here on)}\label{null_conditions}
\end{equation} 
that are satisfied globally in space-time. Under these constraints, the electric and magnetic field lines of a globally null light field evolve like stretchable elastic filaments embedded in a shear-free flow \cite{irvine_linked_2010,robinson_null_1961}.

Such globally null light fields further arise naturally  in  contexts ranging from  radiating electromagnetic waves \cite{bateman_mathematical_1915}, to the geometry of space-time \cite{robinson_null_1961,peres_null_1960,peres_geometrodynamics_1961,geroch_electromagnetism_1966}. 
Their many interesting properties raise a natural and long standing \cite{coll_permanence_1988,mariot_champ_1954-1} 
question: are there  initial conditions on a light field, which guarantee that it will remain null under time evolution?

Our main result (Eq.~(\ref{f_main_result})) is that an initially null electromagnetic field which is analytic everywhere and has a smooth associated flow, stays null forever if and only if the flow is initially shear-free.

The shear-free nature of the flow of null light fields had previously been studied only for globally null light fields \cite{robinson_null_1961,bampi_shear-free_1978} where it was thought to be key to the persistence of knots \cite{irvine_linked_2010,kedia_tying_2013}. However, previous attempts at finding the initial conditions that guarantee nullness had either been inconclusive \cite{coll_permanence_1988,bampi_shear-free_1978} or reached the erroneous conclusion that an initially null light field stays null \cite{mariot_champ_1954-1}.

We begin by clarifying the relation between the null condition and the flow of electromagnetic fields, correcting an error in \cite{irvine_linked_2010} where the flow was assumed to be incompressible. We then extend the previous analysis to establish a mapping between Maxwell's  equations and the Euler equations for a pressureless fluid. We use this mapping to show that an initially null, analytic electromagnetic field is globally null if and only if its flow is initially shear-free, answering a longstanding open question.

We go on to outline methods for constructing initially knotted null, shear-free light fields which would be guaranteed to stay knotted and null. Lastly, we give sufficient conditions for the electric and magnetic field lines of a globally null light field to be tangent to surfaces, i.e. for first integrals to exist, in terms of the Bateman potentials \cite{bateman_mathematical_1915,kedia_tying_2013,hogan_bateman_1984} which are known to exist for all globally null light fields. This allows one to reduce the problem of analyzing the topology of field lines to dimension 2, i.e. the level sets of the first integral.

Our results pave the way for the design of new persistently knotted null light fields, and give tools for studying their topology.

\begin{figure}[!htb]
\includegraphics[width=\columnwidth]{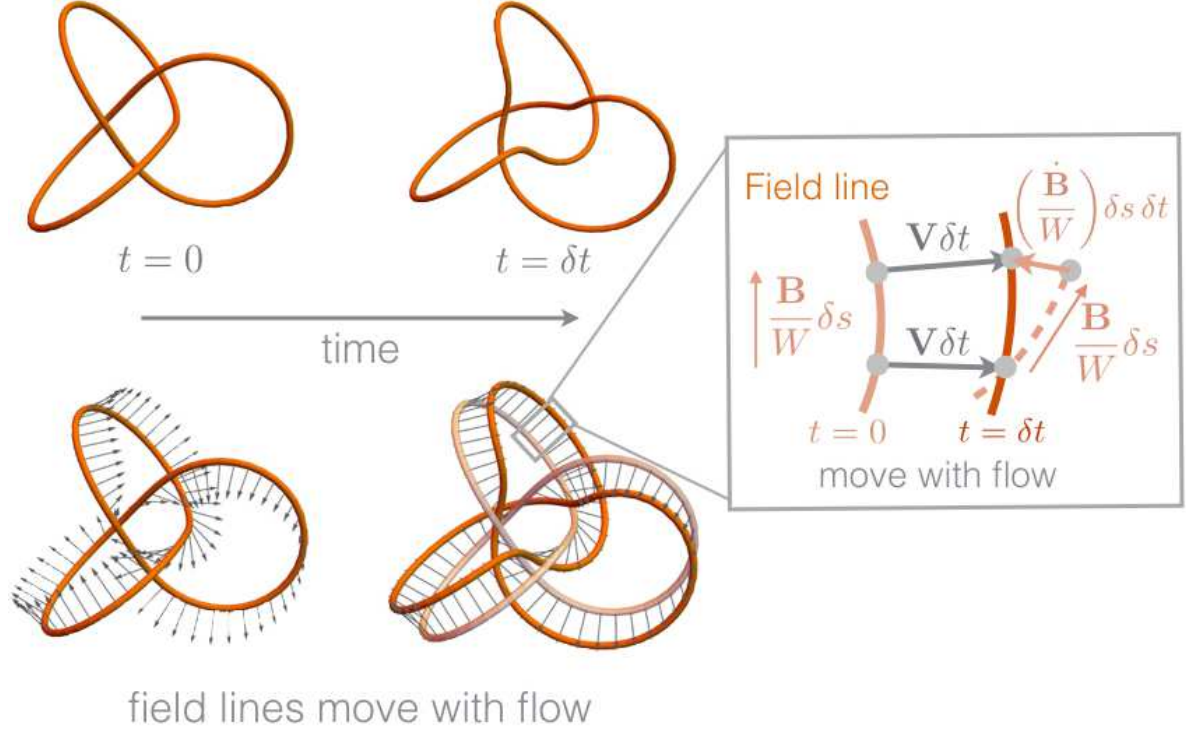}
\caption{Transport of a knotted magnetic field line (orange) along the flow of the Poynting field (grey) in a null light field. The evolution of the magnetic field is such that each magnetic field line moves along the flow of the Poynting field. This is because the field lines of $\mathbf{B}/W$ are transported along $V$, as shown in the inset, and described by Eq.~(\ref{diffeo}). The null light field depicted above is a trefoil knotted field from the knotted null light fields presented in \cite{kedia_tying_2013}.}
\label{field_line_transport}
\end{figure}

\section{Flow of null light fields}\label{sec:hydro_null}

A knotted magnetic (or electric) field line in a globally null light field stays knotted as it evolves, as shown in Fig.~\ref{field_line_transport}. 
The knot deforms smoothly with time, with each point on the knotted field line moving in the direction of the Poynting field at that point.

This smooth deformation of magnetic field lines by a flow  occurs if the magnetic flux through an arbitrary loop co-moving with the flow is preserved~\cite{newcomb_motion_1958,irvine_linked_2010}, i.e. the magnetic (electric) field evolves according to:
\begin{equation}
\partial_t\mathbf{B} = \nabla\times\left( \mathbf{V}\times\mathbf{B} \right) \label{flux_preserving}
\end{equation}
where $\mathbf{V}$ is the flow field.
The null conditions (Eq.~(\ref{null_conditions})) automatically enforce such  \cite{newcomb_motion_1958} a flux-preserving evolution of the magnetic (electric) field along a flow given by the normalized Poynting field $\mathbf{V}=(\mathbf{E}\times\mathbf{B})/W$, where $W=\left( \mathbf{E}^2+\mathbf{B}^2 \right)/2$.

When the flow of the normalized Poynting field $\mathbf{V}$ is incompressible, that is if $\nabla \cdot \mathbf{V}=0$, the flux-preserving evolution of the magnetic (electric) field implies the transport of the magnetic (electric) field lines along the flow of $\mathbf{V}$ \cite{irvine_linked_2010} described by:
\begin{equation}
\partial_t\mathbf{B} + \left(\mathbf{V}\cdot\nabla\right)\mathbf{B} = \left(\mathbf{B}\cdot\nabla\right)\mathbf{V}
\end{equation}
The flow of $\mathbf{V}$ is however not,   incompressible in general. For compressible flows  the lines of the magnetic (electric) field are nonetheless transported since the lines of the normalized magnetic (electric) field $\mathbf{B}/W$ ($\mathbf{E}/W$ respectively) are transported along the flow of $\mathbf{V}$ for any globally null light field, i.e. 
\begin{eqnarray}
\partial_t\left( \frac{\mathbf{B}}{W} \right) + \left(\mathbf{V}\cdot\nabla\right)\left( \frac{\mathbf{B}}{W}\right) = \left( \frac{\mathbf{B}}{W} \cdot\nabla\right) \mathbf{V} \label{diffeo}
\end{eqnarray}
as depicted in Fig.~\ref{field_line_transport}.
Here, we assume that the normalized Poynting field $\mathbf{V}$ is smooth everywhere, which is possible even when the energy density $W$ vanishes as for the knotted null light fields in \cite{kedia_tying_2013}. The transport of the magnetic (electric) field lines along the flow of $\mathbf{V}$ in Eq.~(\ref{diffeo}) is reminiscent of the transport of vortex lines in an inviscid barotropic flow $\mathbf{u}$, given by:
\begin{equation} 
\partial_t\left(\frac{\bomega}{\rho} \right)  + \left(\mathbf{u}\cdot\nabla\right)\left( \frac{\bomega}{\rho} \right) = \left( \frac{\bomega}{\rho} \cdot\nabla \right) \mathbf{u}  \label{omega_transport} 
\end{equation}
where the vorticity field $\bomega$ is $\bomega:=\nabla\times\mathbf{u}$, and $\rho$ is the mass density of the fluid.

\begin{figure}[!htb]
\includegraphics[width=\columnwidth]{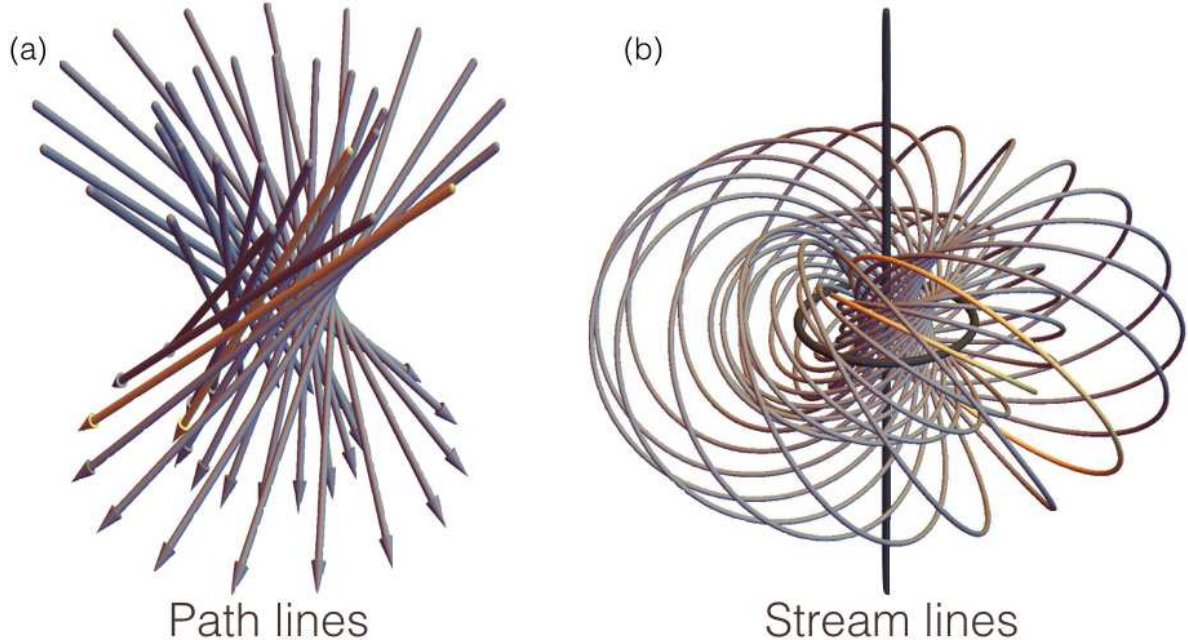}
\caption{Path lines and stream lines of the flow $\mathbf{V}$ for the knotted null light fields presented in \cite{kedia_tying_2013}. The path lines of energy packets are transported by straight line trajectories (as shown by the path lines highlighted in yellow), the geodesics given by $(1,\mathbf{V})$. However, the stream lines of $V$ at an instant in time form closed loops (as shown by the loops highlighted in yellow) such that each loop is linked with every other loop, organized around the z-axis and the unit circle in the x-y plane shown in dark grey, forming a structure known as the Hopf fibration.}
\label{path_stream_null}
\end{figure}

In addition to transporting the electric and magnetic field lines along its flow, the normalized Poynting field $\mathbf{V}$ satisfies the Euler equation for a pressure-less flow:
\begin{equation}
\partial_t \mathbf{V} + \left(\mathbf{V}\cdot\nabla\right)\mathbf{V} = 0 \label{V_euler}
\end{equation}
thus making the analogy with fluid flow complete. The above nonlinear equation arises from a linear system of equations---Maxwell's equations---as a result of the nonlinear null conditions. Under the null conditions, the spatial part of the electromagnetic stress-energy tensor simplifies to that of a pressure-less fluid: $T^{ij}=W\,V^iV^j$, and momentum conservation becomes the pressure-less Euler equation (See \ref{app:appendixA} for a more detailed derivation). 

Furthermore, the vorticity field associated with the flow of the Poynting field $\bOmega:=\nabla\times\mathbf{V}$, is also transported along the flow of the Poynting field:
\begin{equation}
\partial_t\left(\frac{\bOmega}{W}\right) + \left(\mathbf{V}\cdot\nabla\right)\left(\frac{\bOmega}{W} \right) = \left(\frac{\bOmega}{W}\cdot\nabla \right)\mathbf{V} \label{poynting_vorticity}
\end{equation}
The similarities between pressure-less Euler flow and the evolution of a null light field are summarized below in Table~\ref{tab:idealfluidanalogy},
\begin{table}[!htb]
\centering
\begin{tabular}[c]{clc}
\hline
 {\bf Null light field} & $\leftrightarrow$ & {\bf Pressure-less Euler flow} \\
\hline
 Energy conservation  &  & Mass conservation   \T\\
$\partial_t W + \nabla\cdot\left( W\, \mathbf{V}\right) = 0$  & $\leftrightarrow$ & $\partial_t\rho + \nabla\cdot\left( \rho\, \mathbf{u}\right) = 0$\\
  {\footnotesize energy density} $W$ & & {\footnotesize mass density} $\rho$ \B\\
\hline
Momentum conservation & &  Momentum conservation \T \\
$\partial_t\mathbf{V} + \left(\mathbf{V}\cdot\nabla\right)\mathbf{V}=0$ & $\leftrightarrow$ & $\partial_t\mathbf{u} + (\mathbf{u}\cdot\nabla)\mathbf{u} = 0$ \\
$\mathbf{V} := (\mathbf{E}\times\mathbf{B})/W$ & & {\footnotesize fluid velocity} $\mathbf{u}$  \B\\
nonlinear null conditions & & nonlinear Euler equation  \B\\
\hline
Transport of $\mathbf{E}/W,\,\mathbf{B}/W,\, \bOmega/W$  & &  Transport of $\bomega/\rho$ \T\\
$\partial_t\left(\pmb{\Omega}/W \right) = \left[ \,\left(\pmb{\Omega}/W\right)\,,\,\mathbf{V} \,\right]$  & $\leftrightarrow$ & $\partial_t\left(\pmb{\omega}/\rho \right) = \left[ \,\left(\pmb{\omega}/\rho\right)\,,\,\mathbf{u} \,\right]$ \B\\
 $\partial_t\left(\mathbf{B}/W \right) = \left[ \,\left(\mathbf{B}/W\right)\,,\, \mathbf{V}\, \right]$ & & 
\B\\ $\partial_t\left(\mathbf{E}/W \right) = \left[\,\left(\mathbf{E}/W\right)\,,\,\mathbf{V}\,\right]$ & & 
\B\\
$\bOmega := \nabla\times\mathbf{V}$  & & $\bomega := \nabla\times\mathbf{u}$ 
\B\\
\hline
\end{tabular}
\centering{\caption{Flow of null light fields (detailed proofs in \ref{app:appendixA}) \label{tab:idealfluidanalogy}}}
\end{table}
in which we use $\left[ \,\mathbf{X}\,,\,\mathbf{Y}\, \right] = \left( \mathbf{X}\cdot\nabla \right) \mathbf{Y} - \left( \mathbf{Y}\cdot\nabla \right) \mathbf{X}$ to denote the commutator of vector fields. 

In addition to energy and momentum, inviscid barotropic flows have an additional conserved quantity known as helicity~\cite{moffatt_degree_1969} $\mathcal{H} = \int {\rm d}^3x\, \mathbf{u}\cdot\bomega$---a measure of the average linking between the lines of the vorticity field $\bomega$.  An analogous conservation law holds for the helicity density $h_\Omega = \mathbf{V}\cdot\Omega$ associated with the flow of the Poynting field $\mathbf{V}$:
\begin{equation}
\partial_t h_\Omega + \nabla\cdot\left( \mathbf{V}\,h_\Omega \right) = 0 \label{h_omega}
\end{equation}
The electric and magnetic helicity is also conserved for globally null light fields \cite{irvine_linked_2010}.

Remarkably, the analogy we established between null light fields and the flow of an inviscid pressure-less fluid also holds for a local time slice, if the null conditions are satisfied on that time slice. 
This suggests a route to  answer the longstanding open question: when does an electromagnetic field satisfying the null conditions on an initial time slice, satisfy the null conditions for all time? 

\section{When does an initially null light field stay null?}
 Since globally null light fields preserve the topology of the electric and magnetic field lines, the initial conditions that ensure nullness for all space-time also ensure the persistence of knots for all time. Even though null light fields have been studied extensively \cite{mariot_champ_1954,robinson_null_1961,peres_null_1960,peres_geometrodynamics_1961,bampi_shear-free_1978,coll_permanence_1988,geroch_electromagnetism_1966}, these initial conditions have remained elusive. 
Previous attempts at finding the initial conditions that guarantee nullness for all space-time, have either been inconclusive \cite{bampi_shear-free_1978,coll_permanence_1988} or have reached the incorrect conclusion \cite{mariot_champ_1954} that satisfying null conditions on the initial time slice is sufficient to guarantee nullness for all space-time. We  use the analogy between the evolution of null light fields and fluid flow to find the necessary and sufficient initial conditions that ensure nullness for all space-time.

The intuitive ideas that underlie our proof, formalized in  \ref{app:appendixB}, are as follows. We first show that there is a space-time region U of the form $U=\{(t,\mathbf{r})\in \mathbb R^4, 0<t<\psi(\mathbf{r}), \psi\in C^\infty(\mathbb R^3)\}$ where the null condition holds. Nullness for all space time is then ensured by the propagation properties of Maxwell's equations. The evolution of an initially null light field at $t=0$, is akin to the transport of a triplet of mutually perpendicular vectors $\{\mathbf{E},\mathbf{B}, \mathbf{V} \}$ along the flow of $\mathbf{V}$, as in Eqs.~(\ref{diffeo}),(\ref{V_euler}). Preserving the null conditions with time requires preserving the angles between and magnitudes of $\mathbf{E}\,,\,\mathbf{B}$ along the flow of $\mathbf{V}$. For the angle between the electric and magnetic fields $\{ \mathbf{E},\mathbf{B} \}$ and their magnitudes to remain invariant under the flow of $\mathbf{V}$, the flow of $\mathbf{V}$ must be shear-free, i.e. must be free of non-uniform stretching or twisting. An initially null light field preserves nullness when transported along the shear-free flow of its normalized Poynting field $\mathbf{V}$.
Furthermore, a shear-free flow stays shear-free when transported along itself in a flat space-time.

Hence, an initially null light field with an initially shear-free flow of the normalized Poynting field, satisfies the null conditions and preserves the shear-free nature of the flow at later times, establishing the null and shear-free conditions as sufficient initial conditions for a light field to be globally null. 
Furthermore, since the flow of the normalized Poynting field $\mathbf{V}$ of an initially null light field that satisfies the null conditions at later times, is necessarily shear-free \cite{robinson_null_1961},  the null and shear-free conditions are also necessary. 

As a consequence, an initially null electromagnetic field which is analytic everywhere with a smooth normalized Poynting field $\mathbf{V}$, stays null forever, if and only if the flow of $\mathbf{V}$ is initially shear-free, i.e.
\begin{eqnarray}
 &\left\{ \begin{array}{l} \mathbf{E}\cdot\mathbf{B} = 0 \\ \mathbf{E}\cdot\mathbf{E}=\mathbf{B}\cdot\mathbf{B}  \end{array}\; \&\;
\begin{array}{l} (E^i\,E^j - B^i \,B^j)\, \partial_j V_{i} = 0 \\
(  E^i\, B^j + E^j\,B^i )\, \partial_j V_{i} = 0 \end{array} \right\}_{t=0}  \nonumber \\
&\qquad\qquad \Leftrightarrow \left\{ \begin{array}{l} \mathbf{E}\cdot\mathbf{B} = 0 \\ \mathbf{E}\cdot\mathbf{E}=\mathbf{B}\cdot\mathbf{B}  \end{array} \right\}_{\forall \,t} \label{main_result}
\end{eqnarray}

We  further note that a deep connection between conformal foliations and shear-free light rays \cite{baird_cr_2013}, allows the shear-free condition given above  in Eq.~(\ref{main_result}) to be expressed in much simpler terms: 
\begin{eqnarray}
 &\left\{ \mathbf{F}\cdot\mathbf{F}=0\, ,\,
\mathbf{F}\cdot\nabla\times\mathbf{F}=0 \right\}_{t=0}  \Leftrightarrow \left\{ \mathbf{F}\cdot\mathbf{F}=0 \right\}_{\forall \,t} \label{f_main_result}
\end{eqnarray}
where $\mathbf{F}=\mathbf{E}+i\mathbf{B}$ is the Riemann-Silberstein vector, $\mathbf{F}\cdot\mathbf{F}=0$ encodes the null constraints, and $\mathbf{F}\cdot\nabla\times\mathbf{F}=0$ encodes the shear-free condition (see \ref{app:conformal} for details). 

The results presented above could be used in conjunction with methods for the  construction of knotted vector fields \cite{kedia_weaving_2016} to design  null electromagnetic fields with non-trivial topology.

\section{Constructing initially null, shear-free light fields}
In this section, we  describe possible routes for the construction of initially null, shear-free light fields, which are then guaranteed to satisfy the null conditions for all time.

\subsection{Initially knotted null light fields}
A knotted divergence-free vector field can be constructed from a rational map $\psi(\mathbf{r})$ based on Milnor polynomials \cite{dennis_isolated_2010}, as shown in \cite{kedia_weaving_2016,bode_constructing_2016,bode_knotted_2017}:
\begin{equation}
\mathbf{B} = f(\chi,\eta) \nabla\chi\times \nabla\eta = f(\chi,\eta)\frac{\nabla\psi^\ast\times\nabla\psi}{i\,(1+\psi\,\psi^\ast)^2} \label{knotted_B}
\end{equation}
where $\chi(\mathbf{r}) =\left( \psi(\mathbf{r})\,\psi^\ast(\mathbf{r}) \right)/\left(1+\psi(\mathbf{r})\,\psi^\ast(\mathbf{r}) \right)$, $\chi(\mathbf{r}):\mathbb{R}^3\to [0,1]$  and $\eta(\mathbf{r}) = \frac{1}{2 i}\log\left(\psi(\mathbf{r})/\psi^\ast(\mathbf{r}) \right)$, $\eta(\mathbf{r}):\mathbb{R}^3\to [0,2\pi)$. 
To construct an initially null electromagnetic field, a normalized Poynting field perpendicular to the above knotted magnetic field $\mathbf{B}$ can be constructed as follows:
\begin{equation}
\mathbf{V} = \frac{g(\chi,\eta)\nabla\chi + h(\chi,\eta) \nabla\eta}{\vert g(\chi,\eta)\nabla\chi + h(\chi,\eta) \nabla\eta \vert} \label{perp_V}
\end{equation}
so that $\mathbf{V}\cdot\mathbf{V}=1$. 

The above knotted magnetic field $\mathbf{B}$, and the normalized Poynting field $\mathbf{V}$, determine an electric field $\mathbf{E}:=\mathbf{B}\times\mathbf{V}$ which automatically satisfies the initial null conditions.

To define an initially knotted, null and shear-free light field, it is sufficient to solve for the functions $f(\chi,\eta),g(\chi,\eta),h(\chi,\eta)$ such that the electric field $\mathbf{E}$ is divergence-free, and the normalized Poynting field $\mathbf{V}$ is shear-free, i.e.
\begin{eqnarray}
& \nabla\cdot\mathbf{E} = 0 \label{divfree_E} \\
& \mathbf{E}\cdot\nabla\times\mathbf{E} - \mathbf{B}\cdot\nabla\times\mathbf{B} = 0 \label{sf_1}\\
& \mathbf{B}\cdot\nabla\times\mathbf{E} + \mathbf{E}\cdot\nabla\times\mathbf{B} = 0 \label{sf_2}
\end{eqnarray}
The existence of solutions $f,g,h$ to the system of nonlinear PDEs given by Eq.s~(\ref{divfree_E})-(\ref{sf_2}) is a very delicate issue. If there exist functions $f(\chi,\eta),g(\chi,\eta),h(\chi,\eta)$ such that the light field $\{\mathbf{E},\mathbf{B},\mathbf{V}\}$ is analytic, then such a light field will stay knotted, null and shear-free.

\subsection{Initially null, shear-free light fields}
 Alternatively, an initially null, shear-free light field can be constructed from a pair of conjugate functions \cite{nurowski_construction_2010} $f,g$, which are real-valued functions of space such that $\vert \nabla f\vert = \vert \nabla g\vert\,,\, \nabla f\cdot \nabla g = 0$. An initially null, shear-free electromagnetic field $\mathbf{F}=\mathbf{E}+i\mathbf{B}$ can be constructed as $\mathbf{F} = p(\mathbf{x})\,\left( \nabla f(\mathbf{x}) + i\,\nabla g(\mathbf{x})\right)$, where $p(\mathbf{x})$ is a complex-valued function of space chosen to make $\mathbf{F}$ divergence-free. Such an initial electromagnetic field automatically satisfies the null conditions $\mathbf{F}\cdot\mathbf{F}=0$ and the shear-free condition $\mathbf{F}\cdot\nabla\times\mathbf{F}=0$. 

All known persistently knotted light fields \cite{kedia_tying_2013} can be expressed in the above form $\mathbf{F}= p(\mathbf{x}) \, \nabla \left( f(\mathbf{x}) + i\, g(\mathbf{x})\right)$ with $f(\mathbf{x}) + i\,g(\mathbf{x}) = 4(x-i\,y)/(r^2-t^2-1+2\,i\,z)$, and 
\[p(\mathbf{x})= m n \frac{\left( r^2-t^2-1+2\,i\,z \right)^{m+1}\left(2(x-i\,y)\right)^{n-1}}{\left(r^2-(t-i)^2 \right)^{m+n+1}}\] where $r^2=x^2+y^2+z^2$, and $m,n$ are positive integers.

Following Nurowski \cite{nurowski_construction_2010} we can construct all possible pairs of analytic conjugate functions $(f,g)$ as follows. We begin with an arbitrary holomorphic function $F:\mathbb C^2\to\mathbb C$, and find the complex-valued functions $\phi_1(x,y,z)$ and $\phi_2(x,y,z)$ that are solutions to the following algebraic system of equations:
\begin{eqnarray}\label{eqconj}
(x+iy)\phi_1+z\phi_2&=\frac{\partial F(\phi_1,\phi_2)}{\partial\phi_1}\,,\\-(x-iy)\phi_2+z\phi_1&=\frac{\partial F(\phi_1,\phi_2)}{\partial\phi_2}\,.\label{eqconj2}
\end{eqnarray}
Then the pair of conjugate functions $(f,g)$ is given as the real and imaginary parts of the following quadrature
\[
f+ig=C+\int (\phi_1^2-\phi_2^2)\,dx\,,
\]
where $C$ is an arbitrary constant. 
The system of equations~(\ref{eqconj}) and~(\ref{eqconj2}) does not admit, in general, solutions $(\phi_1,\phi_2)$ that are smooth on the whole $\mathbb{R}^3$.   

To understand the structure of the field lines of the null electromagnetic fields constructed via the above methods, we use Bateman's framework \cite{bateman_mathematical_1915,hogan_bateman_1984,kedia_tying_2013} to study their geometry.

\section{Geometry of null electromagnetic fields}
Studying the structure of field lines in knotted light fields is challenging owing to the difficulty of explicitly solving for the field lines everywhere in space. The existence of first integrals i.e. global surfaces such that the field lines are tangent to these surfaces, has made it possible to study the field line structure of knotted light fields in terms of the surfaces they lie on \cite{arrayas_class_2015,irvine_linked_2008,kedia_tying_2013}. We now give sufficient conditions for the existence of such first integrals for a  null light field. 

A method of representing null electromagnetic fields  using complex scalar potentials was given by Bateman \cite{bateman_mathematical_1915}. Hogan \cite{hogan_bateman_1984}, has shown that all null electromagnetic fields can be  expressed in terms of Bateman potentials $(\alpha, \beta)$: complex scalar functions of space-time  which satisfy:
\begin{equation}
{\mathbf\nabla}\alpha\times{\mathbf\nabla}\beta=i\left(\partial_{t}\alpha{\mathbf\nabla}\beta
-\partial_{t}\beta\,{\mathbf\nabla}\alpha\right)\label{alphabeta_def}
\end{equation}
The corresponding null electromagnetic field is given by:
\begin{equation}
\mathbf{F}=  \mathbf{E} + i\mathbf{B} = \nabla\alpha\times\nabla\beta \label{emdef_bateman_basic}
\end{equation}

The design of knotted null electromagnetic fields requires understanding the interplay between the topology of the lines of the electric and magnetic fields, and the complex scalar potentials $(\alpha, \beta)$. 
Unfortunately, describing the topology of the lines of a space-filling vector field is an intricate problem, made all the more challenging by the difficulty of analytically solving for the field lines everywhere in space. However, in the special case that the lines of a vector field are tangent to surfaces, i.e. the vector field has a first integral, the topology of the field lines is encoded in the topology of these surfaces. In the recently discovered family of knotted Maxwell fields \cite{kedia_tying_2013}, the field lines were tangent to knotted tori, and establishing the persistence of these knotted tori helped establish persistent knottedness of the field lines.

We find that the field lines of a null electromagnetic field are tangent to surfaces if the surfaces of constant magnitude of the Bateman potentials $\alpha$ and $\beta$ are parallel, i.e.
\begin{eqnarray}
&\nabla\left( \alpha^\ast \,\alpha \right)\times\nabla\left( \beta^\ast\, \beta \right)=0 \label{mag_condition} \\
&\quad \Rightarrow \mathbf{E}\cdot\nabla\left( \textrm{Im}\{\alpha\beta\}\right) =0 \,,\,\mathbf{B}\cdot\nabla\left( \textrm{Re}\{\alpha\beta\}\right) =0 \label{mag_first_integral}
\end{eqnarray}
where $\alpha^\ast,\beta^\ast$ are complex conjugates of $\alpha,\beta$. The above condition given by Eq.~(\ref{mag_condition}) holds for all known persistently knotted light fields \cite{kedia_tying_2013}. 
Alternatively, if the surfaces of constant phase of $\alpha$ and $\beta$ are parallel, i.e.
\begin{eqnarray}
&\nabla\left( \frac{\alpha}{\alpha^\ast} \right)\times\nabla\left( \frac{\beta}{\beta^\ast} \right)=0 \label{phase_condition} \\
&\quad \Rightarrow \mathbf{E}\cdot\nabla\left( \textrm{Re}\{\alpha\beta\}\right) =0 \,,\,\mathbf{B}\cdot\nabla\left( \textrm{Im}\{\alpha\beta\}\right) =0 \label{phase_first_integral}
\end{eqnarray}
giving surfaces tangent to the electric and magnetic field lines everywhere.

The topology of the field lines of null electromagnetic fields whose Bateman potentials $(\alpha,\beta)$ satisfy either Eq.~(\ref{mag_condition}) or Eq.~(\ref{phase_condition}) can be studied via the topology of the isosurfaces of $\textrm{Re}\{\alpha\beta\}\,,\, \textrm{Im}\{\alpha\beta\}$.

\section{Summary}
Our results firmly establish the connection between the evolution of null electromagnetic fields by Maxwell's equation and transport along flow of the Poynting field. We used this connection to show that an initially null light field stays null if and only if the flow of the normalized Poynting field is initially shear-free. 
Remarkably enough, a surprising connection with the theory of conformal foliations, allows us to express the shear-free condition at t=0 in a very simple way.
This gives a design rule for constructing persistently knotted solutions to Maxwell's equations: satisfying the null conditions and the shear-free condition on a time slice is sufficient to guarantee the persistence of knots. 

We presented ways of constructing initially null light fields that satisfy the shear-free condition. Lastly, we gave sufficient conditions for the lines of null electromagnetic fields to lie on surfaces, enabling the study of the field line structure in terms of surfaces.

\ack
The authors are grateful to  the KITP for
hospitality during the early part of this work. W.T.M.I. and H.K. 
acknowledge support from the National Science
Foundation (NSF) Grant No. DMR-1351506 and the Packard Foundation; D.P.-S. is supported by the ERC Starting Grant 335079 and the ICMAT-Severo Ochoa grant SEV-2015-0554.

\begin{appendix}
\section{Null electromagnetic fields and Euler flows} \label{app:appendixA}

We begin by considering a null electromagnetic field, i.e. an electromagnetic field satisfying $\mathbf{E}\cdot\mathbf{B}=0\,,\,\mathbf{E}\cdot\mathbf{E}=\mathbf{B}\cdot\mathbf{B}$ (in units where $c=1$), and give proofs for the equations stated in Table~\ref{tab:idealfluidanalogy}.

\begin{enumerate}
\item $\partial_t W + \nabla\cdot\left(W\,\mathbf{V} \right) = 0$, where $W=\frac{1}{2}\left( \mathbf{E}^2+\mathbf{B}^2 \right)$ and $\mathbf{V}=\left(\mathbf{E}\times\mathbf{B}\right)/W$.

The above follows from energy conservation for an electromagnetic field in vacuum, and holds for any electromagnetic field \cite{landau_classical_1975}.

\item $\partial_t\mathbf{V} + \left(\mathbf{V}\cdot\nabla\right)\mathbf{V}=0$

The above equation follows from noting that the spatial part of the stress-energy tensor for a null light field simplifies to $T^{ij} = W\,V^iV^j$, and momentum conservation. Momentum conservation for an electromagnetic field in vacuum gives \cite{landau_classical_1975}:
\begin{equation}
\partial_t\left( W\, V_i \right) + \partial_j\left(W\,\delta_{ij} - E_i\,E_j - B_i\,B_j \right) = 0 \label{mom_cons}
\end{equation}

For null electromagnetic fields, the following special property holds $(W\,\delta_{ij} - E_i\,E_j - B_i\,B_j) = W\,V_i\,V_j$. Substituting in Eq.~(\ref{mom_cons}), and using energy conservation, we get
\begin{equation}
\partial_t V_i + V_j \,\partial_j\, V_i = 0 \label{null_geodesic}
\end{equation}
as required.

\item $\partial_t\left(\mathbf{E}/W \right) = \left[\mathbf{E}/W\,,\, \mathbf{V}\right]\,,\, \partial_t\left(\mathbf{B}/W \right) = \left[\mathbf{B}/W\,,\, \mathbf{V}\right]$

For null electromagnetic fields, $\mathbf{V}\times\mathbf{E}=\mathbf{B}\,,\, \mathbf{V}\times\mathbf{B}=-\mathbf{E}$. Substituting, we find that $\partial_t\mathbf{E} = \nabla\times(\mathbf{V}\times\mathbf{E})\,,\, \partial_t\mathbf{B} = \nabla\times(\mathbf{V}\times\mathbf{B})$ follows trivially from Maxwell's equations. Using the definition of the Lie Bracket $\left[ \mathbf{A}\,,\,\mathbf{B}\right]=\left(\mathbf{A}\cdot\nabla\right)\mathbf{B} - \left(\mathbf{B}\cdot\nabla\right)\mathbf{A}$, and energy conservation of electromagnetic fields, the above relation follows.
\end{enumerate}
We note that the above statements and proofs do not assume that the null conditions are satisfied for all times, and hold true on a time slice if the null conditions are satisfied on a time-slice.

We now show the conservation of electric helicity $\mathcal{H}_e = \int\textrm{d}^3x\,\mathbf{C}\cdot\mathbf{E}\,,\, \nabla\times\mathbf{C}=\mathbf{E}$, magnetic helicity $\mathcal{H}_m = \int\textrm{d}^3x\,\mathbf{A}\cdot\mathbf{B}\,,\, \nabla\times\mathbf{A}=\mathbf{B}$, and Poynting helicity $\mathcal{H}_\Omega = \int\textrm{d}^3x\,\mathbf{V}\cdot\Omega\,,\, \nabla\times\mathbf{V}=\Omega$. Since the electric $\mathbf{E}$, magnetic $\mathbf{B}$ and the Poynting vorticity $\Omega$ fields all obey an equation of type $\partial_t\mathbf{Y}=\nabla\times\left( \mathbf{V}\times \mathbf{Y} \right)$, we will show how such an equation implies the conservation of the helicity of a vector field $\mathbf{Y}$, i.e. $\mathcal{H}_Y = \int\textrm{d}^3x\,\mathbf{Z}\cdot\mathbf{Y}\,,\,\nabla\times\mathbf{Z}=\mathbf{Y}$, and the conservation of $\mathcal{H}_e\,,\,\mathcal{H}_m\,,\,\mathcal{H}_\Omega$ follows automatically.

The evolution of $\mathcal{H}_Y$ is given by:
\[ \partial_t \mathcal{H}_Y = \int\textrm{d}^3x\,\partial_t\mathbf{Z}\cdot\mathbf{Y}+\int\textrm{d}^3x\,\mathbf{Z}\cdot\partial_t\mathbf{Y}\]

The evolution of $\mathbf{Y}$ implies an evolution equation for $\mathbf{Z}$:
\begin{eqnarray}
\partial_t\left(\nabla\times\mathbf{Z} \right) &= \nabla\times\left(\mathbf{V}\times\mathbf{Y} \right) \nonumber \\
\nabla\times\left(\partial_t\mathbf{Z} \right) &= \nabla\times\left(\mathbf{V}\times\mathbf{Y} \right) \nonumber \\
\partial_t\mathbf{Z} &= \mathbf{V}\times\mathbf{Y} + \nabla\,f \,, \, (\exists \textrm{ a scalar function } f)\nonumber 
\end{eqnarray}

Substituting for the evolution of $\mathbf{Z}$ in the evolution of $\mathcal{H}_Y$, we get:
\begin{eqnarray}
\partial_t \mathcal{H}_Y &= \int\textrm{d}^3x\,\left( \mathbf{V}\times\mathbf{Y} + \nabla\,f\right)\cdot\mathbf{Y}+ \nonumber \\
&\qquad
\int\textrm{d}^3x\,\mathbf{Z}\cdot\nabla\times\left(\mathbf{V}\times\mathbf{Y}\right) \nonumber \\
&=\int\textrm{d}^3x\, \Big[ \left(\mathbf{V}\times\mathbf{Y} \right)\cdot\mathbf{Y} + \nabla\,f\cdot\mathbf{Y}  + \nonumber \\ 
&\qquad \nabla\cdot\left((\mathbf{V}\times\mathbf{Y})\times\mathbf{Z}\right) + \left(\nabla\times\mathbf{Z} \right)\cdot\left(\mathbf{V}\times\mathbf{Y} \right) \Big]\nonumber \\
&= \int\textrm{d}^3x\,\nabla\cdot\left[ \left( f\,\mathbf{Y}\right) + \left((\mathbf{V}\times\mathbf{Y})\times\mathbf{Z}\right)  \right] = 0 \nonumber
\end{eqnarray}

\section{Shear-free transport, Nullness and Maxwell's equations}\label{app:appendixB}

The proof proceeds by showing that the null and shear-free conditions on the initial time slice are transported along the Poynting field. Using nullness, we then show that the divergence-free property is also transported. Finally using these, we show that $\mathbf{E},\mathbf{B}$ satisfy Maxwell's equations in free space.
By the uniqueness theorems, evolution of the initial electromagnetic field by Maxwell's equations with the given initial conditions should also give null solutions. Thus the shear-free condition is the required initial condition for an initial null electromagnetic field to stay null.

Detailed proofs leading up to Eq.~(\ref{main_result}) are given below.

\begin{lemma}\label{transport_poynting_finite_t} Consider the initial-value problem:
\begin{equation}
\partial_{t}\mathbf{V}+\left( \mathbf{V}\cdot\nabla \right) \mathbf{V} = 0\, , \quad \mathbf{V}_{t=0} = \mathbf{V}_0  \label{lem_eul_geod_re}
\end{equation}
There exists a unique $C^\infty$ solution $\mathbf{V}(\mathbf{r},t)$ in a space-time region $U$, defined as
$U := \left\{ (t,\mathbf{r}) \in \mathbb{R}^4 : 0< t < \psi(\mathbf{r})\, , \, \psi \in C^\infty(\mathbb{R}^3) \right\}\subset \mathbb R^3\times\mathbb R^+$, if $\,\mathbf{V}_0(\mathbf{r})\,  \in C^\infty(\mathbb{R}^3)$. 
\end{lemma}
Eq.~(\ref{lem_eul_geod_re}) is simply the geodesic transport of the initial vector field $\partial_t + \mathbf{V}_0$. Consider points $(0,\mathbf{x}_1)$ lying in the neighborhood $N_0$ of a point $(0,\mathbf{x}_0)$ in the hyperplane $\{t=0\}$. The geodesics in $\mathbb{R}^4$ with initial point $(0,\mathbf{x}_1)\in N_0$ in the direction $(1,\mathbf{V}_0(\mathbf{x}_1))$ are straight lines $(S)$:
\[ (S): \left\{ \begin{array}{rcl}
        t(\lambda) &= &\lambda \\
       \mathbf{x}(\lambda) &= &\mathbf{x}_1 + \lambda \mathbf{V}_0(\mathbf{x}_1) 
\end{array} \right. \,,\quad \mathbf{x}_1 \in N_0 \] 
The solutions to Eq.~(\ref{lem_eul_geod_re}) for $(0,\mathbf{x}_1)\in N_0$ are smooth in time $t\in (0,T)$ provided that the straight lines $(S)$ do not intersect for any $\mathbf{x}_1\in N_0$ and $\lambda \in (0,T)$.

To estimate the time of existence, we consider the intersection of two straight lines with initial points $(0,\mathbf{x}_1)\in N_0$ and $(0,\mathbf{x}_2)\in N_0$, i.e. 
\[ \left\{ \begin{array}{rcl}
    \mathbf{x}_1 + \lambda \mathbf{V}(\mathbf{x}_1) &=  &\mathbf{x}_2 + \lambda \mathbf{V}_0(\mathbf{x}_2)    \\
      \lambda &=  &T 
\end{array} \right.  \] 
\[ \Rightarrow T = \frac{| \mathbf{x}_1 - \mathbf{x}_2 |}{ |\mathbf{V}_0 (\mathbf{x}_1) - \mathbf{V}_0 (\mathbf{x}_2)|} \]

It follows that if $\mathbf{V}_0(\mathbf{x})$ is locally Lipschitz continuous at $\mathbf{x}_0$, there exists a constant $C_{N_0}$, which depends on the neighborhood $N_0$ of $\mathbf{x}_0$, s.t. 
\[ \left\{ \begin{array}{rcl}
   |\mathbf{V}_0(\mathbf{x}_1) - \mathbf{V}_0 (\mathbf{x}_0) |  &\leq  &C_{N_0} | \mathbf{x}_1 - \mathbf{x}_0 |   \\
      \textrm{for any}\; \mathbf{x}_1 \in N_0 &{} &{} 
\end{array} \right.  \] 
\[ \Rightarrow T \geq \frac{1}{C_{N_0}} \]
Hence, on the neighborhood $N_0$, there exists a unique $C^\infty$ solution to eq.(\ref{lem_eul_geod_re}) for time $t\in (0,\frac{1}{C_{N_0}})$. Considering a locally finite covering of $\mathbb{R}^3$ by open sets $N_i$, centred at points $\mathbf{x}_i$, we get local time existence for $t\in (0,\frac{1}{C_{N_i}})$ for each set $N_i$, provided $V_0(\mathbf{x})$ is locally Lipschitz continuous at each point of $\mathbb{R}^3$. One can then easily define a space-time region $U := \left\{ (t,\mathbf{x}) \in \mathbb{R}^4 : 0 < t < \psi(\mathbf{x}) \right\}$ such that $\psi \in C^\infty(\mathbb{R}^3)$, and there exists a $C^\infty$, and unique, solution to Eq.~\ref{lem_eul_geod_re} in the set $U$, as we wanted to prove. Notice that the constants $C_{N_i}$ do not need to be uniform in $i$, and hence the region $U$ may be narrower and narrower.

\begin{theorem}\label{main_theorem}
Let $\tilde{\mathbf{E}}(\mathbf{r},t), \tilde{\mathbf{B}}(\mathbf{r},t),\mathbf{V}(\mathbf{r},t)$ be smooth solutions in $\mathbb{R}^3\times\mathbb{R}^+$ to the initial-value problem :
\begin{eqnarray}
& \partial_t\mathbf{V} + \left(\mathbf{V}\cdot\nabla \right) \mathbf{V} = 0\,,  \label{transport_poynting_e_b_1} \\
& \, \partial_t \tilde{\mathbf{E}} = \left[ \tilde{\mathbf{E}}, \mathbf{V} \right] \,,\,  \partial_t \tilde{\mathbf{B}} = \left[ \tilde{\mathbf{B}}, \mathbf{V} \right] \label{transport_poynting_e_b_2} \\
& \tilde{\mathbf{E}}_{t=0}=\frac{\mathbf{E}_0}{\rho_0}\,,\, \tilde{\mathbf{B}}_{t=0}=\frac{\mathbf{B}_0}{\rho_0}\,,  \label{init_poynting_e_b_1} \\
&\mathbf{V}_{t=0} = \frac{\mathbf{E}_0\times\mathbf{B}_0}{\rho_0}\,,\, \rho_0 := \frac{1}{2}\left( \mathbf{E}_0^2 + \mathbf{B}_0^2 \right) \label{init_poynting_e_b_2}
\end{eqnarray}
with the following initial conditions:
\begin{eqnarray}
& \nabla\cdot\mathbf{E}_0 = \nabla\cdot\mathbf{B}_0=0 \label{div_free}\\
& \mathbf{E}_0 \cdot \mathbf{B}_0 = \mathbf{E}_0 \cdot \mathbf{E}_0 - \mathbf{B}_0\cdot\mathbf{B}_0 = 0 \label{null_0} \\
& (E_0^i B_0^j)g_{0_{ij}}=(E_0^i E_0^j - B_0^j B_0^k )g_{0_{ij}}=0\,,\nonumber \\
& g_{0_{ij}}:= \partial_i V_{0j} + \partial_j V_{0i}\label{shear_free_0}
\end{eqnarray}

Then the fields $\mathbf{E}:= \rho \tilde{\mathbf{E}}\,,\, \mathbf{B} := \rho \tilde{\mathbf{B}},{\textstyle \rho := \frac{1}{\frac{1}{2}\left( \tilde{\mathbf{E}}^2 + \tilde{\mathbf{B}}^2 \right)} }$ are null solutions to Maxwell's equations with the initial conditions:
$\mathbf{E}_{t=0} = \mathbf{E}_0 \,,\, \mathbf{B}_{t=0} = \mathbf{B}_0$ where $\rho$ satisfies the continuity equation: $\partial_t \rho + \nabla\cdot\left( \rho \mathbf{V} \right)=0,$ with the initial condition $\rho_{t=0}=\rho_0\,,\, \mathbf{V}=\frac{\mathbf{E}\times\mathbf{B}}{\rho}$. Moreover, $\rho = \frac{1}{2}\left( \mathbf{E}^2 + \mathbf{B}^2 \right)$.

The converse also holds true, i.e. if $\mathbf{E},\mathbf{B}$ are null solutions to Maxwell's equations with $\mathbf{E}_{t=0} = \mathbf{E}_0 \,,\, \mathbf{B}_{t=0} = \mathbf{B}_0$, then the fields $\tilde{\mathbf{E}}:= \frac{\mathbf{E}}{\rho}\,, \tilde{\mathbf{B}}:= \frac{\mathbf{B}}{\rho}, \mathbf{V} = \frac{\mathbf{E}\times\mathbf{B}}{\rho}$ satisfy Eqs.~(\ref{transport_poynting_e_b_1}),(\ref{transport_poynting_e_b_2}) with the initial conditions given by Eqs.~(\ref{init_poynting_e_b_1}),(\ref{init_poynting_e_b_2}) and Eqs.~(\ref{div_free}),(\ref{null_0}),(\ref{shear_free_0}).

Notice that Eq.~(\ref{shear_free_0}) is simply the shear-free condition at $t=0$ written in Eq.~(\ref{main_result}).
\end{theorem}
To prove the first part of the theorem, we need to prove that if the initial normalized Poynting field $\mathbf{V}_0$ is shear-free, i.e. it satisfies Eq.~(\ref{shear_free_0}), then $\mathbf{E},\mathbf{B}$ generated by solving Eqs.~(\ref{transport_poynting_e_b_1}),(\ref{transport_poynting_e_b_2}) with initial conditions in Eqs.~(\ref{init_poynting_e_b_1}),(\ref{init_poynting_e_b_2}),(\ref{div_free}),(\ref{null_0}),(\ref{shear_free_0}) are null solutions to Maxwell's equations.

Given fields $\mathbf{E}_0(\mathbf{r}),\mathbf{B}_0(\mathbf{r})$ at time $t=0$, we define a tetrad $({k_0}^\mu, {n_0}^\mu, {\bar{m}_0}^\mu, {m_0}^\mu)$:
\begin{eqnarray}
{k_0}^\mu &= \frac{1}{\sqrt{2}}\left(1,\mathbf{V}_0\right)\, ,\, {n_0}^\mu =\frac{1}{\sqrt{2}}\left(1,-\mathbf{V}_0\right) \label{initial_tetrad_ln} \\
{m_0}^\mu &=\left(0,\frac{\mathbf{E}_0+\mathrm{i}\mathbf{B}_0}{\rho_0\,\sqrt{2}}
\right)\, ,\,
{\bar{m}_0}^{\mu} =\left(0,\frac{\mathbf{E}_0-\mathrm{i}\mathbf{B}_0}{\rho_0\,\sqrt{2}}
\right) \label{initial_tetrad_mmbar}
\end{eqnarray}

Solving Eqs.~(\ref{transport_poynting_e_b_1}),(\ref{transport_poynting_e_b_2}) is akin to defining the tetrad $(k^\mu, n^\mu, \bar{m}^\mu, m^\mu)$ for all space-time by parallel transport of $k,n$ and Lie transport of $m$:
\begin{equation}
k^\mu\partial_\mu k^\nu = 0\,,\,  k^\mu\partial_\mu n^\nu = 0\,,\, \left[ k, m \right]^{\mu} = 0 \label{tetrad_evolution}
\end{equation}
with the initial conditions: $\left(k^\mu, n^\mu, \bar{m}^\mu, m^\mu\right)_{t=0}=({k_0}^\mu, {n_0}^\mu, {\bar{m}_0}^\mu, {m_0}^\mu)$ and defining the fields $(\mathbf{E}/\rho,\mathbf{B}/\rho,\mathbf{V})$ by this tetrad as:
\begin{eqnarray}
k^\mu &= \frac{1}{\sqrt{2}}\left(1,\mathbf{V}\right)\,,\, n^\mu =\frac{1}{\sqrt{2}}\left(1,-\mathbf{V}\right) \label{time_dep_V} \\
m^\mu &=\left(0,\frac{\mathbf{E}+\mathrm{i}\mathbf{B}}{\rho\,\sqrt{2}} \right)\,,\,
\bar{m}^{\mu} =\left(0,\frac{\mathbf{E}-\mathrm{i}\mathbf{B}}{\rho\,\sqrt{2}} \right)
\label{time_dep_EB_rho}
\end{eqnarray}
We now show that the fields $\mathbf{E},\mathbf{B}$ are null solutions to Maxwell's equations, related to $\mathbf{V}$ and $\rho$ by: $\mathbf{V}=\mathbf{E}\times\mathbf{B}/\rho$,  and 
$\rho=\frac{1}{2}(\mathbf{E}\cdot\mathbf{E}+\mathbf{B}\cdot\mathbf{B})$, if the initial conditions given by Eqs.~(\ref{div_free}),(\ref{null_0}),(\ref{shear_free_0}) are satisfied.

The sequence of steps followed in the proof is as follows. We begin by showing that the null and shear-free conditions on the initial time slice, i.e. Eqs.~(\ref{null_0})(\ref{shear_free_0}), are transported along $k^\mu$. Using nullness, we then show that the divergence-free condition in Eq.~(\ref{div_free}) is also transported by $k^\mu$. Finally using these, we show that $\mathbf{E},\mathbf{B}$ satisfy Maxwell's equations in free space.

We will use the following notation:
\begin{equation}
k^\mu \partial_\mu= \frac{{\rm D}}{{\rm D}r_k}  \, ;\: m^\mu \partial_\mu = \frac{{\rm D}}{{\rm D}r_m} \label{notation_vector_deriv}
\end{equation}

In this notation, Eq.~(\ref{tetrad_evolution}) can be rewritten as:
\begin{equation}
\frac{{\rm D}}{{\rm D}r_k} k^\nu = 0\,,\, \frac{{\rm D}}{{\rm D}r_k}n^\nu=0\,,\, \frac{{\rm D}}{{\rm D}r_k} m^\mu = \frac{{\rm D}}{{\rm D}r_m}k^\mu \label{diffeo_evolution}
\end{equation}

The commutation relation of the vector fields $k^\mu, m^\mu$ above can also be expressed as:
\begin{equation}
\frac{{\rm D}}{{\rm D}r_k} \frac{{\rm D}}{{\rm D}r_m} = \frac{{\rm D}}{{\rm D}r_m}\frac{{\rm D}}{{\rm D}r_k} \label{km_commutation}
\end{equation}

Note that Eq.~(\ref{diffeo_evolution}) for the evolution of $k^\mu,n^\mu$ along with the initial condition Eq.~(\ref{initial_tetrad_ln}) implies that $k^\mu,n^\mu$ can be written as in Eq.~(\ref{time_dep_V}): $k^\mu = \frac{1}{\sqrt{2}}\left(1,\mathbf{V}\right)\,,\, n^\mu =\frac{1}{\sqrt{2}}\left(1,-\mathbf{V}\right)$.

Also, the tetrad when transported along $k^\mu$, remains a tetrad, i.e.
\begin{equation}
\frac{{\rm D}}{{\rm D}r_k}\{k\cdot k\,,\,k\cdot n\,,\,k\cdot m\,,\,n\cdot n\,,\,n\cdot m \}=0 \label{transport_tetrad}
\end{equation}
as a consequence of Eq.~(\ref{tetrad_evolution}). From the null initial conditions it follows that:
\begin{equation}
k^\mu k_\mu = k^\mu m_\mu = n^\mu n_\mu = n^\mu m_\mu = 0\,,\,k^\mu n_\mu=1  \label{null_tetrad1}
\end{equation}

In conjunction with Eq.~(\ref{time_dep_V}), $k\cdot k = 0 \Rightarrow \mathbf{V}\cdot\mathbf{V}=1$. Also $k\cdot m =0 \Rightarrow \mathbf{V} \propto \mathbf{E}\times\mathbf{B}$. This along with the initial condition Eq.~(\ref{initial_tetrad_ln}),  and $\mathbf{V}\cdot\mathbf{V}=1$ implies:
\begin{equation}
\mathbf{V}=\frac{\mathbf{E}\times\mathbf{B}}{\big\vert \mathbf{E}\times\mathbf{B}\big\vert} \label{V_poynting}
\end{equation}

We define $\rho$ by the equation: $\bar{m}^\mu m_\mu =-1/\rho$ consistent with $\rho=\frac{1}{2}(\mathbf{E}\cdot\mathbf{E}+\mathbf{B}\cdot\mathbf{B})$.

To see how nullness changes along the rays $k^\mu$, we compute
\begin{eqnarray}
&\frac{{\rm D}}{{\rm D}r_k} \left( m^\mu m_\mu \right) = 2m^\mu \frac{{\rm D}}{{\rm D}r_k} m_\mu = 2 m^\mu \frac{{\rm D}}{{\rm D}r_m} k_\mu 
\nonumber \\
&\quad = 2 m^\mu \frac{{\rm D}}{{\rm D}r_m} k_\mu = 2 m^\mu m^\nu \partial_\nu k_\mu = 2 \tilde{\sigma} \label{nullness_transport_raw}
\end{eqnarray}
where $\tilde{\sigma}=m^\mu m^\nu \partial_\nu k_\mu = m^\mu \frac{{\rm D}}{{\rm D}r_m} k_\mu$.

When the tetrad $(k,n,\bar{m},m)$ is null, $\tilde{\sigma}=\sigma/\rho$ where $\vert\sigma\vert$ is the shear of the congruence $k^\mu$. 
The shear-free initial condition implies $\tilde{\sigma}_{t=0}=0$ . To see how $\tilde{\sigma}$ changes along the rays $k^\mu$,
\begin{eqnarray}
\frac{{\rm D}}{{\rm D}r_k} \tilde{\sigma} &= \frac{{\rm D}}{{\rm D}r_k}
\left(m^\mu \frac{{\rm D}}{{\rm D}r_m} k_\mu \right) \nonumber \\
&= \frac{{\rm D}}{{\rm D}r_k} m^\mu \frac{{\rm D}}{{\rm D}r_m} k_\mu  + 
m^\mu \frac{{\rm D}}{{\rm D}r_k} \left( \frac{{\rm D}}{{\rm D}r_m}k_\mu \right) \nonumber \\ 
&= \frac{{\rm D}}{{\rm D}r_k} m^\mu \frac{{\rm D}}{{\rm D}r_m} k_\mu + m^\mu \frac{{\rm D}}{{\rm D}r_m} \left( \frac{{\rm D}}{{\rm D}r_k}k_\mu \right) \: {\rm \left(\because\, Eq.~\ref{km_commutation}\right) } \nonumber \\
&= \frac{{\rm D}}{{\rm D}r_k} m^\mu \frac{{\rm D}}{{\rm D}r_m} k_\mu = \frac{{\rm D}}{{\rm D}r_m} k^\mu \frac{{\rm D}}{{\rm D}r_m} k_\mu \: {\rm \left(\because\, Eq.~\ref{diffeo_evolution}\right) }  \label{shear_first_deriv}
\end{eqnarray}

Calculating the second derivative,
\begin{eqnarray}
\frac{{\rm D}^2}{{\rm D}r_k^2} \tilde{\sigma} 
&= \frac{{\rm D}}{{\rm D}r_k} \left( \frac{{\rm D}}{{\rm D}r_m} k^\mu \frac{{\rm D}}{{\rm D}r_m} k_\mu \right) \nonumber \\
&= \frac{{\rm D}}{{\rm D}r_k} \left( \frac{{\rm D}}{{\rm D}r_m} k^\mu \right) \frac{{\rm D}}{{\rm D}r_m} k_\mu  +\nonumber \\
&\qquad
\frac{{\rm D}}{{\rm D}r_m} k^\mu \frac{{\rm D}}{{\rm D}r_k} \left( \frac{{\rm D}}{{\rm D}r_m} k_\mu \right) \nonumber \\
&= \frac{{\rm D}}{{\rm D}r_m} \left( \frac{{\rm D}}{{\rm D}r_k} k^\mu \right) \frac{{\rm D}}{{\rm D}r_m} k_\mu  + \nonumber \\
&\qquad
\frac{{\rm D}}{{\rm D}r_m} k^\mu \frac{{\rm D}}{{\rm D}r_m} \left( \frac{{\rm D}}{{\rm D}r_k} k_\mu \right) \: {\rm \left(\because \, Eq.~\ref{km_commutation} \right) } \nonumber \\
&= 0 \: {\rm \left(\because \, Eq.~\ref{diffeo_evolution} \right) } \nonumber \\
\therefore \frac{{\rm D}^n}{{\rm D}r_k^n} \tilde{\sigma} &= 0\quad \forall \,n \geq 2,\, n\in \mathbb{N} \label{shear_higher_deriv}
\end{eqnarray}

Thus Eqs.~(\ref{diffeo_evolution}),(\ref{km_commutation}) which are equivalent to Eqs.~(\ref{transport_poynting_e_b_1}),(\ref{transport_poynting_e_b_2}) ensure that the $2^{nd}$ and higher order derivatives of $\tilde{\sigma}$ vanish.
The first derivative of $\tilde{\sigma}$ evaluated in Eq.~(\ref{shear_first_deriv}) can be rewritten as follows:
\begin{eqnarray}
\frac{{\rm D}}{{\rm D}r_k} \tilde{\sigma}
&= \frac{{\rm D}}{{\rm D}r_m} k^\mu \frac{{\rm D}}{{\rm D}r_m} k_\mu = \frac{{\rm D}}{{\rm D}r_m} k^\mu \, \delta_\mu^\nu \,\frac{{\rm D}}{{\rm D}r_m} k_\nu \label{shear_transport_raw}
\end{eqnarray}

Since ${m_0}^\mu {m_0}_\mu=0$,  $(k_0,n_0,\bar{m}_0,m_0)$ form a null tetrad and $\left( \delta_\mu^\nu\right)_{t=0}$ can be rewritten as:
\begin{equation}
\left(\delta_\mu^\nu\right)_{t=0}= {k_0}_\mu {n_0}^\nu + {n_0}_\mu {k_0}^\nu - \rho_0 \left( \bar{m_0}_\mu {m_0}^\nu + {m_0}_\mu \bar{m_0}^\nu \right) \label{delta_null_tetrad}
\end{equation}

The above equation can be verified by contracting with each of the basis vectors $(k_0,n_0,\bar{m}_0,m_0)$.  Substituting (\ref{delta_null_tetrad}) in the (\ref{shear_transport_raw}) above, we get,
\begin{eqnarray}
\left( \frac{{\rm D}}{{\rm D}r_k} \tilde{\sigma} \right)_{t=0} &= \left( \frac{{\rm D}}{{\rm D}r_m} k^\mu \right)_{t=0} \Big( {k_0}_\mu {n_0}^\nu + {n_0}_\mu {k_0}^\nu - \nonumber \\
&\: \rho_0 \left( \bar{m_0}_\mu {m_0}^\nu + {m_0}_\mu \bar{m_0}^\nu \right) \Big) \left( \frac{{\rm D}}{{\rm D}r_m} k_\nu \right)_{t=0} \nonumber\\
&= \Bigg( \frac{{\rm D}}{{\rm D}r_m} k^\mu \Big( {k}_\mu {n}^\nu + {n}_\mu {k}^\nu -  \nonumber \\
&\quad \rho \left( \bar{m}_\mu {m}^\nu + {m}_\mu \bar{m}^\nu \right) \Big)  \frac{{\rm D}}{{\rm D}r_m} k_\nu \Bigg)_{t=0} \nonumber \\
&= \Bigg( \frac{{\rm D}k^\mu}{{\rm D}r_m}  \Bigg( -\rho \Big( \bar{m}_\mu {m}^\nu + \nonumber \\
&\qquad {m}_\mu \bar{m}^\nu \Big) \Bigg)  \frac{{\rm D}k_\nu}{{\rm D}r_m}  \Bigg)_{t=0} {\rm \left(\because\, Eq.~\ref{null_tetrad1}\right) } \nonumber \\
&= -\rho_{0}\Bigg( \left( m^\nu \frac{{\rm D}}{{\rm D}r_m} k_\nu \right) \bar{m}_\mu \frac{{\rm D}}{{\rm D}r_m} k^\mu  + \nonumber \\
&\qquad \qquad 
 \left( m_\mu \frac{{\rm D}}{{\rm D}r_m} k^\mu \right) \bar{m}^\nu \frac{{\rm D}}{{\rm D}r_m} k_\nu   \Bigg)_{t=0}
\nonumber \\
&= -\rho_{0} \left( \tilde{\sigma}\,  \bar{m}_\mu \frac{{\rm D}}{{\rm D}r_m} k^\mu + \tilde{\sigma}\,
\bar{m}^\nu \frac{{\rm D}}{{\rm D}r_m} k_\nu  \right)_{t=0} \nonumber \\
&= -2 \tilde{\sigma}_{t=0} \left( \rho  \,\bar{m}_\mu \frac{{\rm D}}{{\rm D}r_m} k^\mu \right)_{t=0} \nonumber \\
&= 0 
\label{shear_transport}
\end{eqnarray}
Eq.~(\ref{shear_transport}) gives:
\begin{equation}
\frac{{\rm D}^n}{{\rm D}r_k^n}\left( m^\mu m_\mu \right)\Big\vert_{t=0} = 0\,,\, \frac{{\rm D}^n}{{\rm D}r_k^n} \tilde{\sigma}\Big\vert_{t=0} = 0, \forall \,n \geq 1 \label{nullness_shear_deriv}
\end{equation}
The above result along with the initial conditions: $\left(m^\mu m_\mu\right)_{t=0} =0$, and $\tilde{\sigma}_{t=0} = 0$, implies that the null and shear-free conditions are transported along $k^\mu$, and hold true for all space-time:
\begin{equation}
m^\mu m_\mu = 0\,,\, \tilde{\sigma}=0  \label{null_shearfree_transport}
\end{equation}

Thus, the geodetic null congruence $k^\mu$ given by the normalized Poynting field $\mathbf{V}$ is shear-free and the fields $\mathbf{E},\mathbf{B}$ defined by Eqs.~(\ref{time_dep_EB_rho}),(\ref{time_dep_V}) are null. Thus $\big\vert\mathbf{E}\times\mathbf{B}\big\vert=\rho$ and Eq.~(\ref{V_poynting}) can be rewritten as: 
\begin{equation}
\mathbf{V}=\frac{\mathbf{E}\times\mathbf{B}}{\rho} \label{V_EcrossB_rho}
\end{equation}

Since ${m}^\mu {m}_\mu=0$,  $(k,n,\bar{m},m)$ form a null tetrad and $\delta_\mu^\nu$ can be rewritten as:
\begin{equation}
\delta_\mu^\nu= {k}_\mu {n}^\nu + {n}_\mu {k}^\nu - \rho \left( \bar{m}_\mu {m}^\nu + {m}_\mu \bar{m}^\nu \right) \label{delta_general}
\end{equation}

We now show that these fields satisfy Maxwell's equations in free space. Calculating the evolution of $1/\rho$ along the null GSF 
$k^\mu$,
\begin{eqnarray*}
\frac{{\rm D}}{{\rm D}r_k} \left( \frac{1}{\rho} \right) &= -\frac{{\rm D}}{{\rm D}r_k} \left( m^\mu \bar{m}_\mu \right) = -m^\mu \frac{{\rm D}}{{\rm D}r_k} \bar{m}_\mu - \bar{m}^\mu \frac{{\rm D}}{{\rm D}r_k} m_\mu \\
&= -m^\mu \bar{m}^\nu \partial_\nu k_\mu - \bar{m}^\mu m^\nu \partial_\nu k_\mu \nonumber \\
&= -\partial_\nu k^\mu \left( m_\mu \bar{m}^\nu + \bar{m}_\mu m^\nu \right)  \\
&= \frac{1}{\rho}\partial_\nu k^\mu \left( -\rho m_\mu \bar{m}^\nu - \rho \bar{m}_\mu m^\nu \right) \\
&= \frac{1}{\rho}\partial_\nu k^\mu \Big( k_\mu n^\nu + n_\mu k^\nu -   \rho m_\mu \bar{m}^\nu - \rho \bar{m}_\mu m^\nu \Big) \nonumber \\
&\qquad \qquad {\rm \left(using\: Eq.~(\ref{null_tetrad1})\right)} \\
&= \frac{1}{\rho} \partial_\nu k^\mu \delta_\mu^\nu = \frac{1}{\rho} \partial_\nu k^\nu \quad {\rm \left(using\: Eq.~(\ref{delta_general})\right)}
\end{eqnarray*}

Substituting for $k^\mu$ from Eq.~(\ref{time_dep_V}), we get the energy conservation equation:
\begin{equation}
\partial_t\, \rho + \nabla\cdot\left( \rho \mathbf{V} \right)=0
\label{rho_time_evol}
\end{equation}

Rewriting Eq.~(\ref{time_dep_EB_rho}) in terms of the Riemann-Silberstein vector $\mathbf{F}=\mathbf{E}+\mathrm{i}\mathbf{B}$,
\begin{equation}
\left( \frac{\mathbf{F}}{\rho} \right)_t = \left[ \mathbf{F}/\rho,\mathbf{V} \right] \label{RS_diffeo}
\end{equation}

Using Eq.~(\ref{rho_time_evol}) along with Eq.~(\ref{RS_diffeo}), 
\begin{eqnarray}
\mathbf{F}_t + \left[ \mathbf{V},\mathbf{F} \right] &= - \big( \nabla\cdot\mathbf{V} \big) \mathbf{F} \nonumber \\
\Rightarrow \mathbf{F}_t+\nabla\times\left(\mathbf{F}\times\mathbf{V}\right) &= -\mathbf{V} \big(\nabla\cdot\mathbf{F} \big) \nonumber \\
\Rightarrow \mathbf{F}_t + \mathrm{i}\nabla\times\mathbf{F} &= -\mathbf{V} \big( \nabla\cdot\mathbf{F}\big) \label{maxwell_2ndset_raw}
\end{eqnarray}

Taking the divergence of Eq.~(\ref{maxwell_2ndset_raw}), 
\begin{eqnarray}
&\left( \nabla\cdot\mathbf{F}\right)_t = -\big(\nabla\cdot\mathbf{V}\big) \big( \nabla\cdot\mathbf{F}\big)-\mathbf{V}\cdot\nabla \big( \nabla\cdot\mathbf{F}\big) \nonumber \\
\Rightarrow &\left(\partial_t +\mathbf{V}\cdot\nabla \right) \big(\nabla \cdot\mathbf{F} \big) = - \big(\nabla\cdot\mathbf{V} \big)\big(\nabla\cdot\mathbf{F} \big) \nonumber \\
\Rightarrow &\frac{{\rm D}}{{\rm D}r_k} \left(\nabla \cdot\mathbf{F} \right) = - \left(\nabla\cdot\mathbf{F} \right) \nabla\cdot\mathbf{V}
\label{divergence_transport}
\end{eqnarray}

Since $\left(\nabla\cdot\mathbf{F}\right)_{t=0}=0$, from Eq.~(\ref{divergence_transport}) above, $\nabla\cdot\mathbf{F}=0$ for all space-time. Thus $\mathbf{E},\mathbf{B}$ are divergence-free for all space-time, satisfying one pair of Maxwell's equations in free space.

Substituting $\nabla\cdot\mathbf{F}=0$ in Eq.~(\ref{maxwell_2ndset_raw}) above, we find that $\mathbf{F}$ also satisfies the other pair of Maxwell's equations: $\mathbf{F}_t + \mathrm{i}\nabla\times\mathbf{F} = 0$. 

Thus the fields $\mathbf{E},\mathbf{B}$ are null solutions to Maxwell's equations in free space.

To see that the converse is true, i.e. a null Maxwell field satisfies the shear-free condition, note that Eq.~(\ref{nullness_transport_raw}) implies that 
\[ \frac{D}{Dr_k}\left( m^\mu m_\mu \right) = 2 \tilde{\sigma} = \frac{2\sigma}{\rho}  \]
Since the electromagnetic field is null,  $m^\mu m_\mu = 0$, which implies $\sigma = 0$, in particular $\sigma_{t=0}=0$, which is equivalent to the condition in Eq.~(\ref{shear_free_0}).

Using Lemma~\ref{transport_poynting_finite_t} and Theorem~\ref{main_theorem} we can finally prove Eq.~(\ref{main_result}), i.e. that null initial conditions are null forever if and only if the shear-free condition is satisfied at $t=0$. This result is presented in the following corollary: 

\begin{corollary}: Let $\mathbf{E},\mathbf{B}$ be solutions to Maxwell's equations in free space with the initial conditions $\mathbf{E}_{t=0}=\mathbf{E}_0\,,\,\mathbf{B}_{t=0}=\mathbf{B}_0$, where $\mathbf{E}_0\,,\,\mathbf{B}_0$ are real analytic, i.e. $C^\omega$ vector fields in $\mathbb{R}^3$, and the initial normalized Poynting field $\mathbf{V}_0$ is well-defined everywhere. The electromagnetic field $\mathbf{E},\mathbf{B}$ is null if and only if $\mathbf{E}_0,\mathbf{B}_0$ satisfy the initial conditions given by Eqs.~(\ref{div_free}),(\ref{null_0}),(\ref{shear_free_0}).
\end{corollary}

In Theorem~\ref{main_theorem}, we proved that if the electromagnetic field $\mathbf{E},\mathbf{B}$ is null, then the initial fields $\mathbf{E}_0, \mathbf{B}_0$ satisfy the conditions given by Eqs.~(\ref{div_free}),(\ref{null_0}),(\ref{shear_free_0}). 

Accordingly, to prove the corollary, we prove that the solution $\mathbf{E},\mathbf{B}$ to Maxwell's equations with initial conditions $\mathbf{E}_{t=0}=\mathbf{E}_0\,,\,\mathbf{B}_{t=0}=\mathbf{B}_0$ where $\mathbf{E}_0,\mathbf{B}_0$  satisfy the conditions given by Eqs.~(\ref{div_free}),(\ref{null_0}),(\ref{shear_free_0}), is a null electromagnetic field. 

Define $\mathbf{V}_0 := \frac{\mathbf{E}_0\times\mathbf{B}_0}{\rho_0}\,,\,\rho_0 := \frac{1}{2}\left( \mathbf{E}_0^2+\mathbf{B}_0^2 \right)$ and consider the initial value problem:
\[ \left\{ \begin{array}{l}
    \mathbf{V}_t +\left( \mathbf{V}\cdot\nabla \right) \mathbf{V} =  0   \\
      \mathbf{V}_{t=0} = \mathbf{V}_0  
\end{array} \right.  \]
Since $\mathbf{V}_0$ is $C^\infty$, it is locally Lipschitz continuous. By Lemma (\ref{transport_poynting_finite_t}), there exists a solution to this initial-value problem, which is $C^\infty$, in the space-time region $U := \Big\{ (t,\mathbf{r}) \in \mathbb{R}^4 : 0< t < \psi(\mathbf{r})\, , \, \psi \in C^\infty(\mathbb{R}^3) \Big\}$.

Observe that, since $\mathbf{V}_{t=0}^2=\mathbf{V}_0^2=1$, we have that $\mathbf{V}^2(\mathbf{x},t)=1$ in $U$, so that $\mathbf{V}(\mathbf{x},t)$ defines a local flow $\phi_t$ in $U$. Accordingly there exist $C^\infty$ solutions to the transport equations:
\[ \partial_t \tilde{\mathbf{E}} = \left[ \tilde{\mathbf{E}}, \mathbf{V} \right]\,,\, \partial_t \tilde{\mathbf{B}} = \left[ \tilde{\mathbf{B}}, \mathbf{V} \right] \]
$\Rightarrow \tilde{\mathbf{E}}(\mathbf{x},t)=\phi_{t\ast}\tilde{\mathbf{E}}_0(\mathbf{x})$ and $\tilde{\mathbf{B}}(\mathbf{x},t)=\phi_{t\ast}\tilde{\mathbf{B}}_0(\mathbf{x})$, where 
$\tilde{\mathbf{E}}_0=\frac{\mathbf{E}_0}{\rho_0}\,,\, \tilde{\mathbf{B}}_0=\frac{\mathbf{B}_0}{\rho_0}$ and $\phi_t$ is the non-autonomous flow of $\mathbf{V}(\mathbf{x},t)$ with $\phi_0(\mathbf{x})=\mathbf{x}$.

Therefore, we have $C^\infty$ solutions in $U$ to the initial value problem:
\[  \left\{   \begin{array}{c}
	\partial_t \mathbf{V} + \left( \mathbf{V}\cdot\nabla \right) \mathbf{V}  = 0 \\
	\partial_t \tilde{\mathbf{E}} = \big[ \tilde{\mathbf{E}},\mathbf{V} \big]  \\
	\partial_t \tilde{\mathbf{B}} = \big[ \tilde{\mathbf{B}},\mathbf{V} \big] \\
	\tilde{\mathbf{E}}_{t=0} = \tilde{\mathbf{E}}_0 \, , \, \tilde{\mathbf{B}}_{t=0} = \tilde{\mathbf{B}}_0 \,,\, \mathbf{V}_{t=0} = \mathbf{V}_0	   
\end{array}         \right.       \]
such that conditions in Eqns~(\ref{div_free}),(\ref{null_0}) and (\ref{shear_free_0}) of Theorem \ref{main_theorem} hold. 

Applying the theorem, we conclude that defining $\rho = \frac{1}{\frac{1}{2}\left(\tilde{\mathbf{E}}^2 + \tilde{\mathbf{B}}^2\right)} $ and $\mathbf{E}=\rho\tilde{\mathbf{E}}\,,\,\mathbf{B}=\rho\tilde{\mathbf{B}}$, the fields $\mathbf{E}(\mathbf{x},t),\mathbf{B}(\mathbf{x},t)$ solve Maxwell's equations in $U$, and satisfy the null conditions in $U$. 

By the existence properties of Maxwell's equations, $\mathbf{E}(t,\mathbf{x})$ and $\mathbf{B}(t,\mathbf{x})$ can be globally propagated, so that they provide $C^\infty$ global solutions to the initial-value problem: 
\[ \left\{ \begin{array}{l}
\partial_t\mathbf{E} = \nabla\times\mathbf{B} \\
\partial_t\mathbf{B} = - \nabla\times\mathbf{E} \\
\mathbf{E}_{t=0}=\mathbf{E}_0 \,,\, \mathbf{B}_{t=0} = \mathbf{B}_0
\end{array} \right. \]

Since $\mathbf{E}_0\,,\,\mathbf{B}_0$ are real-analytic, the solution to Maxwell's equations (which is unique), is real-analytic both in space and time.
Therefore since $\mathbf{E}\cdot\mathbf{B}\,,\,\mathbf{E}\cdot\mathbf{E}-\mathbf{B}\cdot\mathbf{B}$ vanish in $U$, they vanish everywhere, as we wanted to prove.

\section{The shear-free condition and conformal foliations}\label{app:conformal}
In this Appendix we prove Eq.~(\ref{f_main_result}), i.e. that an initial electromagnetic field $\mathbf{F}_0=\mathbf{E}_0+i\mathbf{B}_0$ gives rise to a global null solution of Maxwell's equations if and only if $\mathbf{F}_0\cdot\mathbf{F}_0=0$ and $\mathbf{F}_0\cdot \nabla\times \mathbf{F}_0=0$. Indeed, the shear-free condition for the initial Poynting field $\mathbf{V}_0$, cf. Eq.~(\ref{main_result}), can be written, after symmetrizing, as:
\begin{eqnarray}
& (E_0^iE_0^j - B_0^iB_0^j )(\partial_iV_{0j}+\partial_jV_{0i})=0\label{app_sf_1}\\ &E_0^iB_0^j(\partial_iV_{0j}+\partial_jV_{0i})=0\label{app_sf_2}
\end{eqnarray}
Let us now show that this is equivalent to $V_0$ defining a conformal foliation. As shown in \cite{baird_cr_2013}, the field $V_0$ defines a conformal foliation if and only if the following conditions are satisfied:
\begin{eqnarray}
&\mathbf{V}_0\cdot[(\mathbf{E}_0\cdot\nabla)\mathbf{B}_0+(\mathbf{B}_0\cdot\nabla)\mathbf{E}_0]=0 \label{cf_1}\\ &\mathbf{V}_0\cdot[(\mathbf{E}_0\cdot\nabla)\mathbf{E}_0-(\mathbf{B}_0\cdot\nabla)\mathbf{B}_0]=0\label{cf_2}
\end{eqnarray}
where $\mathbf{E}_0\,,\,\mathbf{B}_0$ satisfy the null conditions, i.e. $\mathbf{F}_0\cdot\mathbf{F}_0=0$. 

It is easy to show that the shear-free conditions are equivalent to the conformal foliation conditions. Using $V_{0i}B_0^i = V_{0i} E_0^i =0$, and $\partial_j\left( V_{0i}B_0^i\right) = \partial_j\left(V_{0i} E_0^i\right)=0$, Eq.~(\ref{app_sf_1}) becomes:
\begin{eqnarray*}
&V_{0j}\left( (E_0^i\,\partial_i) E_0^j - (B_0^i\,\partial_i)B_0^j \right) = 0 
\end{eqnarray*}
i.e. equivalent to Eq.~(\ref{cf_1}), and Eq.~(\ref{app_sf_2}) becomes:
\begin{eqnarray*}
&V_{0j}\left( (E_0^i \partial_i) B_0^j + (B_0^i \partial_i) E_0^j \right) = 0
\end{eqnarray*}
i.e. equivalent to Eq.~(\ref{cf_2}). 

Finally, using that ${\bf E_0}$ and ${\bf B_0}$ satisfy the null conditions, and the following identity for two arbitrary vector fields $X,Y$: 
\begin{eqnarray*}
&\nabla(\mathbf{X}\cdot \mathbf{Y})= \\
&\qquad (\mathbf{X}\cdot\nabla)\mathbf{Y}+(\mathbf{Y}\cdot\nabla)\mathbf{X}+\mathbf{X}\times\nabla\times \mathbf{Y}+\mathbf{Y}\times\nabla\times\mathbf{X}
\end{eqnarray*}
we find that the conformal foliation condition is equivalent to the following:
\begin{eqnarray}
&\mathbf{E}_0\cdot\nabla\times \mathbf{E}_0-\mathbf{B}_0\cdot\nabla\times \mathbf{B}_0=0\\&\mathbf{E}_0\cdot \nabla\times\mathbf{B}_0+\mathbf{B}_0\cdot\nabla\times\mathbf{E}_0=0\,.
\end{eqnarray}
i.e. $\mathbf{F}_0\cdot\nabla\times \mathbf{F}_0=0$. We have then proved that the LHS in Eqs.~(\ref{main_result}) and~(\ref{f_main_result}) are equivalent, so the RHS of these equations are equivalent as well, as we wanted to prove.

\section{First integrals of null electromagnetic fields}\label{app:firstint}
In this appendix we give a proof of the sufficient conditions, i.e. Eqs.~(\ref{mag_first_integral}),(\ref{phase_first_integral}), for the existence of first integrals for null electromagnetic fields expressed in terms of the Bateman complex potentials $\{\alpha,\beta \}$.

Let us denote the magnitude and phase of $\alpha$ by $r_\alpha,\theta_\alpha$  and that of $\beta$ by $r_\beta,\theta_\beta$ so that $\alpha=r_\alpha\exp\left(\mathrm{i}\theta_\alpha\right)\,,\,\beta=r_\beta\exp \left( \mathrm{i}\theta_\beta \right)$. 
Then,
\begin{eqnarray}
\nabla\alpha &= \alpha \left( \mathrm{i}\nabla\theta_{\alpha} +\frac{1}{r_\alpha}\nabla r_\alpha \right) \nonumber \\
\nabla\beta &= \beta \left( \mathrm{i}\nabla\theta_{\beta} +\frac{1}{r_\beta}\nabla r_\beta \right) \nonumber
\end{eqnarray}

Then the electric field is given by:
\begin{eqnarray}
\mathbf{E} &=\mathrm{Re}\left\{ \nabla\alpha\times\nabla\beta \right\} \nonumber\\
&=  \mathrm{Re}\left\{ \alpha\beta \right\}  \left(\frac{\nabla r_\alpha}{r_\alpha}\times \frac{\nabla r_\beta}{r_\beta} - \nabla\theta_\alpha\times \nabla\theta_\beta \right) - \nonumber \\ 
&\qquad \mathrm{Im} \left\{ \alpha\beta \right\} \left(\frac{\nabla r_\alpha}{r_\alpha}\times \nabla\theta_\beta - \frac{\nabla r_\beta}{r_\beta}\times \nabla\theta_\alpha \right) \nonumber
\end{eqnarray}

Similarly the magnetic field is given by:
\begin{eqnarray}
\mathbf{B} &=\mathrm{Im}\left\{ \nabla\alpha\times\nabla\beta \right\} \nonumber\\
&=  \mathrm{Re}\left\{ \alpha\beta \right\}\left(\frac{\nabla r_\alpha}{r_\alpha}\times \nabla\theta_\beta - \frac{\nabla r_\beta}{r_\beta}\times \nabla\theta_\alpha \right)  + \nonumber \\ 
&\qquad \mathrm{Im} \left\{ \alpha\beta \right\} \left(\frac{\nabla r_\alpha}{r_\alpha}\times \frac{\nabla r_\beta}{r_\beta} - \nabla\theta_\alpha\times \nabla\theta_\beta \right) \nonumber
\end{eqnarray}

It is easy to check that $\mathrm{Im}\left\{\nabla\left( \alpha\beta \right) \right\}\,,\,\mathrm{Re}\left\{\nabla\left( \alpha\beta \right) \right\}$ can be written as:
\begin{eqnarray}
 \mathrm{Im}\left\{\nabla\left( \alpha\beta \right) \right\} &= \mathrm{Re}\left\{ \alpha\beta\right\}\left( \nabla\theta_\alpha +\nabla\theta_\beta \right) + \nonumber \\ 
 &\qquad \mathrm{Im}\left\{ \alpha\beta\right\} \left( \frac{\nabla r_\alpha}{r_\alpha} + \frac{\nabla r_\beta}{r_\beta} \right)\nonumber \\
\mathrm{Re}\left\{\nabla\left( \alpha\beta \right) \right\} &= \mathrm{Re}\left\{ \alpha\beta\right\} \left( \frac{\nabla r_\alpha}{r_\alpha} + \frac{\nabla r_\beta}{r_\beta} \right) - \nonumber \\ 
&\qquad \mathrm{Im}\left\{ \alpha\beta\right\} \left( \nabla\theta_\alpha +\nabla\theta_\beta \right) \nonumber
\end{eqnarray}

Using the above relations, we find:
\begin{eqnarray}
\mathbf{E}\cdot\mathrm{Im}\left\{\nabla\left( \alpha\beta \right) \right\} &= r_\alpha r_\beta\left(\nabla r_\alpha \times \nabla r_\beta \right)\cdot\left( \nabla\theta_\alpha +\nabla\theta_\beta \right) \nonumber \\
\mathbf{E}\cdot\mathrm{Re}\left\{\nabla\left( \alpha\beta \right) \right\} &= - r_\alpha r_\beta\left(\nabla \theta_\alpha \times \nabla \theta_\beta \right)\cdot\nabla\left( r_\alpha\, r_\beta \right) \nonumber \\
\mathbf{B}\cdot\mathrm{Re}\left\{\nabla\left( \alpha\beta \right) \right\} &= -r_\alpha r_\beta \left(\nabla r_\alpha \times \nabla r_\beta \right)\cdot\left( \nabla\theta_\alpha +\nabla\theta_\beta \right) \nonumber \\
\mathbf{B}\cdot\mathrm{Im}\left\{\nabla\left( \alpha\beta \right) \right\} &= -r_\alpha r_\beta \left(\nabla\theta_\alpha  \times \nabla\theta_\beta \right)\cdot\ \nabla \left(r_\alpha\,r_\beta\right)  \nonumber
\end{eqnarray}
Hence, if $\nabla r_\alpha\times \nabla r_\beta =0$, $\mathbf{E}\cdot\mathrm{Im}\left\{\nabla\left( \alpha\beta \right) \right\} = 0\,,\,\mathbf{B}\cdot\mathrm{Re}\left\{\nabla\left( \alpha\beta \right) \right\}= 0$. Similarly, if $\nabla \theta_\alpha\times \nabla \theta_\beta =0$,
$\mathbf{E}\cdot\mathrm{Re}\left\{\nabla\left( \alpha\beta \right) \right\} = 0\,,\, \mathbf{B}\cdot\mathrm{Im}\left\{\nabla\left( \alpha\beta \right) \right\}= 0$.

\end{appendix}

\section*{References}
\bibliography{flow_refs}

\providecommand{\newblock}{}
\begin{thebibliography}{10}
\expandafter\ifx\csname url\endcsname\relax
  \def\url#1{{\tt #1}}\fi
\expandafter\ifx\csname urlprefix\endcsname\relax\def\urlprefix{URL }\fi
\providecommand{\eprint}[2][]{\url{#2}}

\bibitem{faraday_lines_1852}
Faraday M 1852 {\em On {Lines} of {Magnetic} {Force}: {Their} {Definite}
  {Character} and {Their} {Distribution} {Within} a {Magnet} and {Through}
  {Space}\/} (Royal Society)

\bibitem{kelvin_vortex_1869}
Kelvin W~T~B 1869 {\em On vortex motion\/} (Royal Society of Edinburgh)

\bibitem{kleckner_creation_2013}
Kleckner D and Irvine W~T~M 2013 {\em Nat. Phys.\/} {\bf 9} 253--258

\bibitem{tkalec_reconfigurable_2011}
Tkalec U, Ravnik M, Copar S, Zumer S and Musevic I 2011 {\em Science\/} {\bf
  333} 62--65

\bibitem{dennis_isolated_2010}
Dennis M~R, King R~P, Jack B, O'Holleran K and Padgett M~J 2010 {\em Nat.
  Phys.\/} {\bf 6} 118--121

\bibitem{hall_tying_2016}
Hall D~S, Ray M~W, Tiurev K, Ruokokoski E, Gheorghe A~H and Mottonen M 2016
  {\em Nat Phys\/} {\bf 12} 478--483

\bibitem{enciso_knots_2012}
Enciso A and Peralta-Salas D 2012 {\em Ann. of Math.\/} {\bf 175} 345--367

\bibitem{enciso_existence_2015}
Enciso A and Peralta-Salas D 2015 {\em Acta Math.\/} {\bf 214} 61--134

\bibitem{faddeev_stable_1997}
Faddeev L and Niemi A~J 1997 {\em Nature\/} {\bf 387} 58--61

\bibitem{houghton_rational_1998}
Houghton C~J, Manton N~S and Sutcliffe P~M 1998 {\em Nucl. Phys. B\/} {\bf 510}
  507--537

\bibitem{manton_topological_2004}
Manton N and M~Sutcliffe P 2004 {\em Topological solitons\/} (Cambridge
  University Press)

\bibitem{aratyn_exact_1999}
Aratyn H, Ferreira L~A and Zimerman A~H 1999 {\em Phys. Rev. Lett.\/} {\bf 83}
  1723--1726

\bibitem{ranada_knotted_1990}
Ra\~nada A~F 1990 {\em J. Phys. A: Math. Gen.\/} {\bf 23} L815--L820

\bibitem{ranada_topological_1989}
Ra\~nada A~F 1989 {\em Lett. Math. Phys.\/} {\bf 18} 97--106

\bibitem{ranada_topological_1992}
Ra\~nada A~F 1992 {\em J. Phys. A: Math. Gen.\/} {\bf 25} 1621--1641

\bibitem{irvine_linked_2008}
Irvine W~T~M and Bouwmeester D 2008 {\em Nat. Phys.\/} {\bf 4} 716--720

\bibitem{urbantke_hopf_2003}
Urbantke H~K 2003 {\em J. Geom. Phys.\/} {\bf 46} 125--150

\bibitem{trautman_solutions_1977}
Trautman A 1977 {\em Int. J. Theor. Phys.\/} {\bf 16} 561--565

\bibitem{arrayas_knots_2017}
Arrayas M, Bouwmeester D and Trueba J~L 2017 {\em Phys. Rep.\/} {\bf 667} 1--61

\bibitem{kedia_tying_2013}
Kedia H, Bialynicki-Birula I, Peralta-Salas D and Irvine W~T~M 2013 {\em Phys.
  Rev. Lett.\/} {\bf 111} 150404

\bibitem{robinson_null_1961}
Robinson I 1961 {\em J. Math. Phys.\/} {\bf 2} 290

\bibitem{irvine_linked_2010}
Irvine W~T~M 2010 {\em J. Phys. A: Math. Theor.\/} {\bf 43} 385203

\bibitem{bateman_mathematical_1915}
Bateman H 1915 {\em The {Mathematical} {Analysis} of {Electrical} and {Optical}
  {Wave}-{Motion}\/} (Dover Publications Inc.)

\bibitem{peres_null_1960}
Peres A 1960 {\em Phys. Rev.\/} {\bf 118} 1105--1110

\bibitem{peres_geometrodynamics_1961}
Peres A 1961 {\em Ann. Phys.\/} {\bf 14} 419 -- 439

\bibitem{geroch_electromagnetism_1966}
Geroch R~P 1966 {\em Ann. Phys.\/} {\bf 36} 147--187

\bibitem{coll_permanence_1988}
Coll B and Ferrando J~J 1988 {\em Gen. Rel. Grav.\/} {\bf 20} 51--64

\bibitem{mariot_champ_1954-1}
Mariot L 1954 {\em C. R. Acad. Sci. Paris\/} {\bf 239} 1189--1190

\bibitem{bampi_shear-free_1978}
Bampi F 1978 {\em Gen. Rel. Grav.\/} {\bf 9} 779--782

\bibitem{hogan_bateman_1984}
Hogan P 1984 {\em Proc. R. Soc. Lond. A\/} {\bf 396} 199--204

\bibitem{newcomb_motion_1958}
Newcomb W~A 1958 {\em Ann. Phys.\/} {\bf 3} 347--385

\bibitem{moffatt_degree_1969}
Moffatt H~K 1969 {\em J. Fluid Mech.\/} {\bf 35} 117--129

\bibitem{mariot_champ_1954}
Mariot L 1954 {\em C. R. Acad. Sci. Paris\/} {\bf 238} 2055--2056

\bibitem{baird_cr_2013}
Baird P and Eastwood M 2013 {\em Ann. Glob. Anal. Geom.\/} {\bf 44} 73--90

\bibitem{kedia_weaving_2016}
Kedia H, Foster D, Dennis M~R and Irvine W~T 2016 {\em Phys. Rev. Lett.\/} {\bf
  117} 274501

\bibitem{bode_constructing_2016}
Bode B and Dennis M~R 2016 {\em arXiv:1612.06328\/}

\bibitem{bode_knotted_2017}
Bode B, Dennis M~R, Foster D and King R~P 2017 {\em Proc. R. Soc. A\/} {\bf
  473} 20160829

\bibitem{nurowski_construction_2010}
Nurowski P 2010 {\em Ann. Glob. Anal. Geom.\/} {\bf 37} 321--326

\bibitem{arrayas_class_2015}
Arrayas M and Trueba J~L 2015 {\em J. Phys. A: Math. Theor.\/} {\bf 48} 025203

\bibitem{landau_classical_1975}
Landau L and Lifshitz E 1975 {\em The {Classical} {Theory} of {Fields}\/}
  Course of theoretical physics, {Vol}. 2 (Butterworth-Heinemann)

\end{thebibliography}

\end{document}